\documentstyle[12pt,epsf]{article}   
\newlength{\dinwidth}
\newlength{\dinmargin}
\def\lsim{\mathrel{\rlap{\lower4pt\hbox{\hskip1pt$\sim$}}
    \raise1pt\hbox{$<$}}}                
\def\gsim{\mathrel{\rlap{\lower4pt\hbox{\hskip1pt$\sim$}}
    \raise1pt\hbox{$>$}}}                
\begin{document}

\newpage

\vspace*{1cm}
\begin{center}  \begin{Large} \begin{bf}
Experimental constraints on coefficients of $\alpha_S$-expansion
of  Gottfried sum rule \\
\end{bf}
\end{Large}
  \vspace*{5mm}
  \begin{large}
A. V. Sidorov$^a$, M.V. Tokarev$^{b}$\\
  \end{large}

\vskip 0.5cm

$^a$Bogoliubov Laboratory of Theoretical Physics,\\
Joint Institute for Nuclear Research,\\
141980 Dubna, Moscow Region, Russia\\

\vskip 0.2cm
$^b$Laboratory of High Energies,\\
Joint Institute for Nuclear Research,\\
141980 Dubna, Moscow Region, Russia\\

\end{center}

\begin{abstract}
The $x$- and $Q^2$-dependences of the Gottfried sum rule
$S_G(x,Q^2)$ based on the  experimental data on
proton and deuteron structure functions  are studied.
The dependence of $S_G(x,Q^2)$  on $Q^2$   for low $x$
points to flavour asymmetry  of sea quark distributions.
 For the first time, the coefficients $c_1, c_2$ of the expansion of
$S_G(x,Q^2)$  in  $\alpha_S/\pi$  up to second order are evaluated through
the phenomenological analysis of NMC and H1 data.
It is found that $c_1$ is negative while $c_2$
is positive. The obtained result  is in disagreement with
QCD predictions for these coefficients.
We  suggest  measuring the low $x$-dependence of the
$F_2^D$ deuteron structure function  at HERA in order to study
 the  $Q^2$-dependence of the  Gottfried sum rule.
\end{abstract}

\newpage

{\section {Introduction}}

The experimental data on the proton and neutron structure functions (SF)
 are of great
 interest for verification of the theory of strong interaction, QCD.
The relevant information  can be used to  extract  the spin-dependent
 and spin-independent parton distributions, to estimate  nonperturbative
effects,  to verify  nucleon models and  sum rules such as the
Gottfried \cite{KG67}, Bjorken \cite{JB66}, Ellis-Jaffe
\cite{JE74}, Gross-Llewellyn Smith \cite{GLS}  and Adler\cite{ADLER}
ones.

 New data on the deuteron  SF
$F_2^D$  obtained  at CERN, SLAC and Fermilab
 \cite{NMC91}-\cite{E665}  stimulated great interest in
  theoretical studies of the deuteron structure.

The deuteron  is an excellent neutron target
and therefore the neutron  structure functions $F_2^n, g_1^n$
 were usually extracted  from the experimentally
known proton and deuteron structure functions.

The extraction procedure  of  the neutron SF
 from deuteron and proton data is ambiguous and,
therefore, the estimate of nuclear effects  in the deuteron
is extremely important not only  to obtain new information
on $F_2^n$  but  also to verify   deuteron models
and to perform a common QCD analysis of experimental data.

The  Gottfried sum rule was verified by the  NMC Collaboration
\cite{NMC91} and the value of $S_G=0.240\pm 0.016$ was found to be below
the  parton model prediction.   To study the sum rule in detail,
the experimental data on the deuteron structure function $F_2^D$ at low $x$
and  large $Q^2$  are necessary. Such measurements are possible
to perform at HERA \cite{DUREN}.

In the present paper, the phenomenological analysis of $x$ and
$Q^2$ dependences of the $S_G(x,Q^2)$ Gottfried sum rule
is based on the NMC \cite{NMC92,NMC95}, H1 \cite{H1}
 and ZEUS \cite{ZEUS} parametrization of the proton
structure function $F_2^p(x,Q^2)$.
It is shown that the available  experimental data on $F_2^p$ and $F_2^D$
allow one to estimate  the $\alpha_S$ correction to $S_G(Q^2)$.
The expansion coefficients of $S_G(Q^2)$
up to order $Q(\alpha^2)$  are estimated. It is found that
the obtained results are in disagreement with the QCD predictions.
To clarify the discrepancy, it was proposed to measure the
$F_2^D(x,Q^2)$ deuteron structure function  at low $x$ and high
$Q^2$ in order to extract precisely the coefficients
of $\alpha_S$-expansion of the Gottfried sum rule.

\vskip 0.5cm

{\section {Deep-Inelastic Scattering on Deuteron}}

 The  cross section of deep - inelastic lepton - deuteron scattering
in the one - photon approximation is expressed via
 the imaginary part  of the forward scattering amplitude
 of the virtual photon or $W$-boson on the deuteron - $W_{\mu \nu}^D$.
The latter  is related to the
deuteron spin-dependent - $g_{1,2}^D(\nu ,Q^2)$ and
spin-independent -  $F_{1,2,3}^{D}(\nu ,Q^2)$
 structure functions as  follows
  $$W_{\mu \nu}^D = -(g_{\mu \nu} - q_{\mu}q_{\nu}/q^2)\cdot F_1^D
  + (p_{\mu}-q_{\mu}(pq)/q^2)(p_{\nu}-q_{\nu}(pq)/q^2)\cdot
 F_2^D/\nu $$
\begin{equation}
 {+ i{\epsilon }_{\mu \nu \alpha \beta} q^{\alpha}
  \{s^{\beta}  g_1^D/\nu
   + [s^{\beta}(qp)-p^{\beta}(sq)]M^{-1} g_2^D/{\nu}^2\}
 + i{\epsilon }_{\mu \nu \alpha \beta}\ q^{\alpha}
   p^{\beta}\cdot  F_3^D/\nu.}
\end{equation}
Here $q, p$ are momenta of the photon and deuteron; $M$ is the deuteron mass;
$\nu=(pq)$; the 4-vector $s_{\alpha}$ describes the deuteron spin.
 The
symmetric part of the deuteron tensor $W_{\mu \nu}^D$  can be
written as $W_{\mu \nu}^D = W_{\mu \nu}^{\alpha \beta}\cdot
 \rho_{\alpha \beta}^{(S)}$

\begin{equation}
W_{\mu \nu}^{\alpha \beta} = \int \frac{d^4 k}{(2\pi)^4i}
\delta (m^2-k^2) {\theta (k_0)} {\theta (p_{+}-k_{+})}\ Sp \{w_{\mu \nu}^N
\cdot \bar {\psi}^{\alpha}(k_1)\cdot (m+\hat k)\cdot
\psi^{\beta}(k_1)\}.
\end{equation}
Here the $\theta $-function and light-cone variables - $(k_{\pm},
 k_{\bot})$ are used.  The  tensor $\rho_{\alpha \beta}^{(S)}$ is the
 symmetric part of the deuteron  polarization  density matrix.  The
 antisymmetric part of the deuteron tensor $W_{\mu \nu}^D$  is
expressed in a form similar to (2). The procedure to construct
the  relativistic  deuteron wave function (RDWF)
${\psi}_{\alpha}$  was proposed and RDWF was obtained in \cite{TOK91}.

\vskip 0.5cm

{\section {Deuteron and Neutron Structure Functions}}

The deuteron SF $F_2^D$ in the light-cone variables
is expressed as follows
\begin{equation}
F_2^D (\alpha , Q^2) = \int_{\alpha}^{1} dx {\ }{d^2}k_{\bot}\ p(x,k_{\bot})
\cdot  F_2^N(\alpha /x, Q^2).
\end{equation}
The nucleon SF $F_2^N=(F_2^p+F_2^n)/2$ is defined by the proton and
neutron ones.
The function $p(x,k_{\bot})$ describes the probability that
 the active nucleon carries away the fraction of  the deuteron momentum
 $x = k_{1+}/p_{+}$
 and the transverse momentum $k_{\bot}$ in the infinite momentum frame.
 It is expressed via  the RDWF

\begin{equation} p(x,k_{\bot}) \propto  Sp\{ \bar
{\psi}^{\alpha}(k_1)\cdot (m+\hat k_1)\cdot \psi^{\beta}(k_1)\cdot \hat
q /\nu \cdot\rho_{\alpha \beta}^{(S)} \}.
\end{equation}

The nuclear effect in the deuteron is described by
the ratio $R_F^{D/N}=F_2^D /F_2^N$~. It was shown in \cite{BRAUN94}
 that   the effect of relativistic Fermi motion  grows
 with $x$  and
the  ratio   $R_F^{D/N}$ reaches 6\%  at $x \simeq 0.7$.  The
dependence of the ratio $R_F^{D/N}$  on $x$ resembles the nuclear EMC
effect and  it is practically independent
of $Q^2$.  Using the universal behaviour of the ratio it is possible to
extract the neutron SF $F_2^n$

\begin{equation} F_2^n (x,Q^2) = 2\cdot
{[R_F^{D/N}(x)]}^{-1} \cdot F_2^D (x,Q^2) - F_2^p (x,Q^2).
\end{equation}

We would like to note  that  for a reliable estimate of
 other  contributions
 to $F_2^D$ from nuclear effects such as nuclear  shadowing, meson
exchanges  etc., the  data at low $x$
 such as the E665 data \cite{E665}  but with  smaller experimental
errors are required.

{\section {Gottfried Sum Rule}}

The extracted  neutron SF can be used to verify the
Gottfried sum rule \cite{KG67}:
\begin{equation}
\int _{0}^{1} [F_2^p (x)-F_2^n (x)]  {dx}/{x} =  {1}/{3}.
\end{equation}

The Gottfried integral  as a function
of $x$ and $Q^2$  is defined as follows
\begin{equation}
S_G (x,Q^2) = \int _{x}^{1} [F_2^p (y,Q^2)-F_2^n (y,Q^2)]
{dy}/{y}.
\end{equation}

To verify the sum rule, not only the $x$-dependence
of SF's also $Q^2$-dependence
in a wide kinematical range are necessary.  The realistic comparison
of the experimental results with theoretical predictions based on QCD
is more argued at high $Q^2$.

As has been reported in \cite{NMC91}, the value of $S_G$
at $Q^2=4.0\ (GeV/c)^2$ obtained
 from the measurements of $F_2^D$ and $F_2^p$ is considerably below the
value of the naive quark-parton model equal to
1/3: $S_G = 0.240 \pm 0.016$.  This result in the parton model is
usually interpreted as the violation of the isospin symmetric sea.
As will be shown later, the asymmetry violation  demonstrates the
strong $Q^2$-dependence.

The QCD corrections of order $O(\alpha_s)$  \cite{Yndur} and
 $O(\alpha_s^2)$  \cite{KKPS} are
estimated for the Gottfried sum rule in the case of
 flavor-symmetric sea $\bar{u}=\bar {d}$:

\begin{equation}
S_G(Q^2)=\frac{1}{3}\cdot (1+\bar c_1\cdot (\alpha_S/\pi)
 + \bar c_2\cdot (\alpha_S/\pi)^2).
\end{equation}

The coefficients $\bar c_1, \bar c_2$ are equal to
0.036,~0.72 for $n_f=3$ and 0.038,~0.55 for $n_f=4$, respectively.
Thus, the coefficients $\bar c_1,~\bar c_2$   are found to be positive and
relatively small, and as mentioned in \cite{KKPS}, the QCD corrections
cannot explain the deviation of the theoretical prediction
from the experimental result of the NMC collaboration without the
assumption that the light quark sea is flavor asymmetric.

\vskip 0.5cm

{\section { Procedure to Extract Neutron
 Structure Function }}

The method to extract  $F_2^n(x,Q^2)$  from proton and deuteron
experimental data was proposed  and realized in \cite{BRAUN94}.

The procedure includes the items

\begin{itemize}
\item experimental data on the ratio $R_F^{D/p}=F_2^D/F_2^p$
and  structure  functions $F_2^p$,\  $F_2^D$

\item the relativistic deuteron model \cite{TOK91}

\item the choice of the parametrization of the $F_2^{n}$
neutron structure function and the determination
of free parameters to describe the ratio $R_F^{D/p}=F_2^D/F_2^p$

\item the  comparison of the absolute values of the experimental and
 theoretical structure function  $F_2^{D}(x,Q^2)$

\item the determination of  the ratio $R_F^{D/N}=F_2^D/F_2^N$
 describing  the nuclear effect in the deuteron

\item the  extraction of the  neutron structure function
>from experimental data using the formula
$$ F_2^n(x,Q^2)= 2\cdot [R_F^{D/N}]^{-1}\cdot F_2^D(x,Q^2)- F_2^p(x,Q^2)$$
\end{itemize}

In \cite{BRAUN94} the NMC data \cite{NMC91,NMC92} on the
ratio $R_F^{D/p}=F_2^D / F_2^p$,
 $F_2^p$  and  the relativistic deuteron model  were used
to extract the neutron SF $F_2^n$.
It was shown that the calculated results for  ratio $R_F^{D/p}$
and   $F_2^{D}(x,Q^2)$
are in good agreement with
the  available experimental data \cite{NMC91,NMC92,SLAC92,BCDMS89}.

Thus,  the conclusion was made  that the extraction
procedure proposed  for $F_2^n(x,Q^2)$ is self-consistent
 because  it  provides a good description
of higher statistics  experimental data
on the ratio $R_F^{D/p}$ and $F_2^D$ over a wide kinematic
range of $x$ and $Q^2$.

We would like to emphasize that the nuclear effect of Fermi motion was
only  used  in \cite{BRAUN94}. The shadowing effect \cite{Badelek}
should be also included
into the procedure  if the  deuteron data in the low $x$- and $Q^2$-  range
 are used ( for example, E665 data). In that case, the factor
$R^{D/N}$ should be corrected at low $x$.

{\section {Results and Discussion}}

Figure 1(a,c,e) shows the  dependence of the Gottfried integral
$S_G(x,Q^2)$ on $x$ and $Q^2$. The parametrization of the
proton structure function $F_2^p(x,Q^2)$ for NMC and H1 data are taken
from \cite{NMC92,NMC95,H1}. The NMC parametrization was used in
\cite{BRAUN94}  for calculating
the                deuteron structure function $F_2^D(x,Q^2)$ and
 a good agreement with
experimental data  SLAC, BCDMS, NMC was obtained both
 for a low and high $x$ range.
It is assumed \cite{BRAUN94} that  the ratio
$f(x)=F_2^n(x,Q^2)/F_2^p(x,Q^2)\rightarrow 1$ as $x \rightarrow 0$.
The parametrization of the $f(x)$ function was obtained  and will be used
later  for the determination  of the neutron SF
from the  H1 and ZEUS  proton structure function parametrization.

One can see from Figure 1(a,c,e) that the $x$- and $Q^2$-dependences
of $S_G(x,Q^2)$ for the parametrizations  are similar.
We would like to note that there is the  crossover  point $x_0$
( in particular, $x_0\sim 0.01 $ for the H1 parametrization)
separating  two ranges: one  - with decreasing   $S_G(x,Q^2)$ and
the other - with increasing $S_G(x,Q^2)$  with $Q^2$, respectively.
A similar dependence  of the  $S_{GLS}(x,Q^2)$
 Gross-Llewellyn Smith integral
on $Q^2$   with the  crossover  point $x_0=\simeq 10^{-2}$ is predicted
in \cite{SID}. Note that the main part of the $S_G(x,Q^2)$ integral
is given by integration over the interval of relatively large $x$
( $10^{-2}< x< 1.$)  and determines  the first  term in expansion (8).

Figure 1(b,d,f) shows the  dependence of the derivative
$dS_G(x,Q^2)/d\alpha_S$ on $x$ and $Q^2$.
The 3-loop perturbative QCD expression of
$\alpha_S$  with $n_f=4$ is used.
The derivative grows up to $x\simeq 0.1-0.2$  then
decreases  and changes in sign for $Q^2>100~GeV^2$ for
NMC and even for smaller $Q^2$ for H1 parametrization.
It should be stressed that {\it {the
negative value of the derivative
$dS_G(x,Q^2)/d\alpha_S$ is due to the small $x$ contribution
to the  Gottfried integral.}}

Figure 2 shows the  dependence of $S_G(Q^2)$ on
$\alpha_S/\pi$ at $x=10^{-3}$  in (7).
 The $\alpha_S$-dependence could be parametrized by the parabola:
$$S_G(Q^2)=S_0\cdot (1+c_1\cdot (\alpha_S/\pi)
 + c_2\cdot (\alpha_S/\pi)^2).$$

The values of $S_0, c_1, c_2$ are presented in Table 1.
They could be considered as $\alpha_S$ corrections to the
Gottfried sum rule.
One can see from Figure 2 that the derivative $dS_G(Q^2)/d\alpha_S$
is negative
at $\alpha_S/\pi<0.05$ for all parametrizations $F_2^p$, and the
H1 curve crosses the NMC one in the range $Q^2=5-10~(GeV/c)^2$.

\vskip 1.cm

\begin{center}
{Table 1.  The  coefficients of the
$S_G(Q^2)$ Gottfried integral expansion in $\alpha_S/\pi$.}\\[0.5cm]

\begin{tabular}{||c|c|c|c||} \hline \hline
                     & $S_0$  &     $c_1$   &  $c_1$    \\ \hline \hline
 NMC92\ \cite{NMC92} & 0.246  & -7.06       &  53.2     \\  \hline
 NMC95\ \cite{NMC95} & 0.210  & -4.87       &  44.2     \\  \hline
 H1\  \ \ \ \cite{H1} & 0.271  & -6.04       &  23.7     \\  \hline \hline
 average value       & 0.242$\pm$ 0.21 & -6.00$\pm$ 0.74  & 40.4$\pm$ 11.1     \\  \hline \hline
\end{tabular}

\end{center}

\vskip 0.5cm

Despite  different kinematical regions of NMC
and H1 experiments used for fits of experimental data
one can see from Table 1 the reasonable quantitative  and good
qualitative agreement between the corresponding coefficients
 $S_0,~c_1,~c_2$ for
parametrizations of the proton SF under consideration.
Note that the  $F_2^p$ parametrization  of H1  is combined with the data from
NMC and BCDMS  experiments  and  a smooth transition between different
$F_2^p$ data is obtained.  The kinematic range covers almost four orders
 of magnitude in $x$ and $Q^2$.

The average values for the   coefficients $r_i\  (i=S_0, c_1, c_2)$
are calculated by $<r>=(\sum_i r_{i})/n$ and
 $<\Delta r>=(\sum_i |<r>- r_{i}|)/n$  and  presented in Table 1.
The errors obtained could be considered as  a crude  estimation of
systematic uncertainties.

It should be noted that the  parametrization of the ZEUS data \cite{ZEUS}
for $F_2^p$ provides qualitatively the same results:
$S_0=0.383$, $c_1=-12.9$ and $c_2=76.2$.

We would like to emphasize that the values of coefficients obtained from
the phenomenological analysis of experimental data are
essentially different from the theoretical QCD predictions
 for $\bar c_1$ and $\bar c_2$.  The coefficient $c_1$ is found to be
negative, in contrast to $\bar c_1$. Both
$c_1,~c_2$ are many times larger in the absolute value than
$\bar c_1,~\bar c_2$.

{\section{ Conclusions}}

The analysis of $Q^2$-dependence of the Gottfried sum rule
based on experimental data on the proton $F_2^p$ and deuteron
$F_2^D$ structure functions  in the framework of
the covariant approach in the light-cone variables
and relativistic deuteron model was performed:

\begin{itemize}
\item  the procedure  to extract the neutron SF
  $F_2^{n}(x,Q^2)$ is described  and  used to analyze  the
parametrizations of
 {NMC, BCDMS, SLAC, H1, ZEUS} data on  $F_2^{D}$ and  $F_2^{p}$ SF's

\item   the increase of the  $S_G(Q^2)$  Gottfried sum rule
         for small $\alpha_S(Q^2)$ is
 a general feature for  the   NMC92, NMC95, H1
 parametrizations (see Fig.2).  This behaviour is connected with
the {\it negative value of the first order $\alpha_S$ correction} to the
GSR

\item it is shown that the results  obtained for   $c_1$ and $c_2$
      are in disagreement  with the calculation made
      in the framework of QCD  assuming the flavour symmetry
      of sea quarks $(\bar u = \bar d)$

\item  the measurements of the deuteron  structure function $F_2^D$
       at HERA, extraction the neutron SF
       $F_2^n$  and verification of the $Q^2$-dependence of $S_G(x,Q^2)$
       at low $x$  are necessary  to determine the
        $\alpha_S$-corrections to GSR more reliably.

\end{itemize}

\begin{center}
{\large \bf Acknowledgement}
\end{center}

The authors would like to thank  Yu.A.Panebratsev
for his support of our work.
This work was partially supported by Grants of the Russian
Foundation for Fundamental Research under   No. 95-02-05061 and No.
96-02-18897. One of us (M.V.T.) are very grateful to
R.Klanner and M.Klein for support to take part  in the Workshop
"Future at HERA".

\newpage

\begin{center}
\vskip -1.cm
\parbox{7cm}{\epsfxsize=7.cm \epsfysize=7.cm \epsfbox[5 5 500 500]
{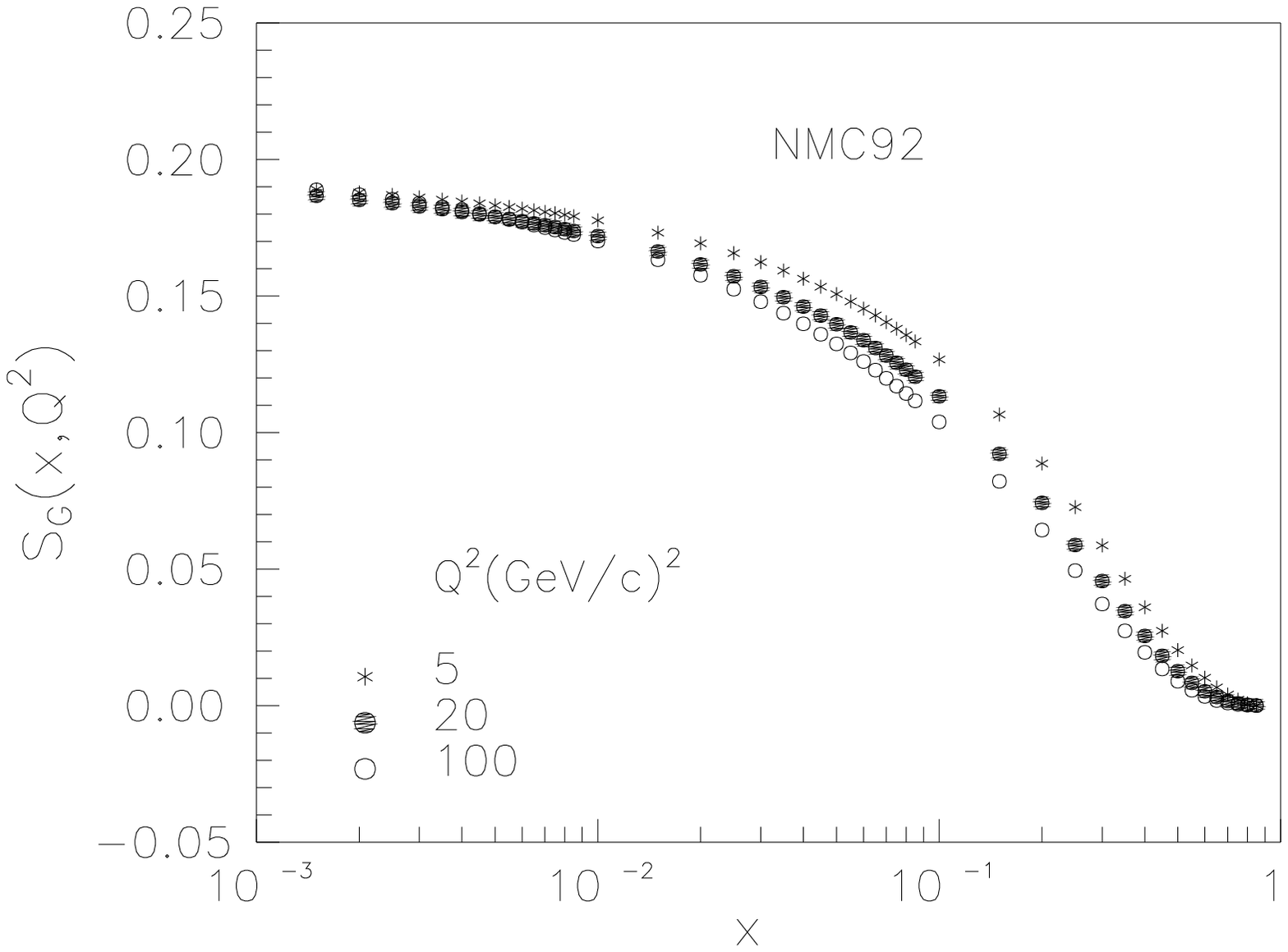}{}}
\parbox{7cm}{\epsfxsize=7.cm \epsfysize=7.cm \epsfbox[35 5 535 500]
{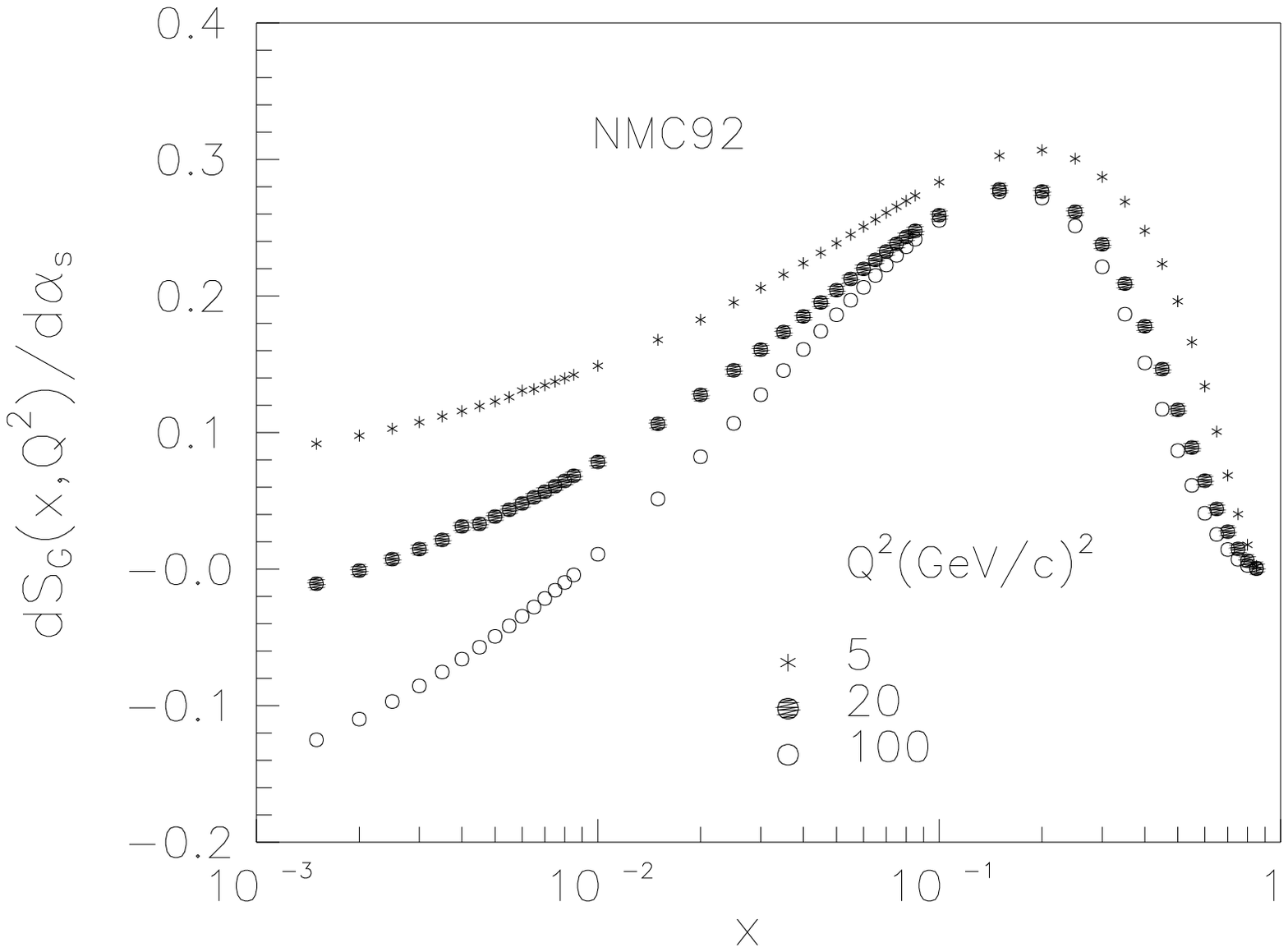}{}}
\end{center}

\begin{center}
\hspace*{0cm} a) \hspace*{7cm} b)
\end{center}

\begin{center}
\parbox{7cm}{\epsfxsize=7.cm \epsfysize=7.cm \epsfbox[5 5 500 500]
{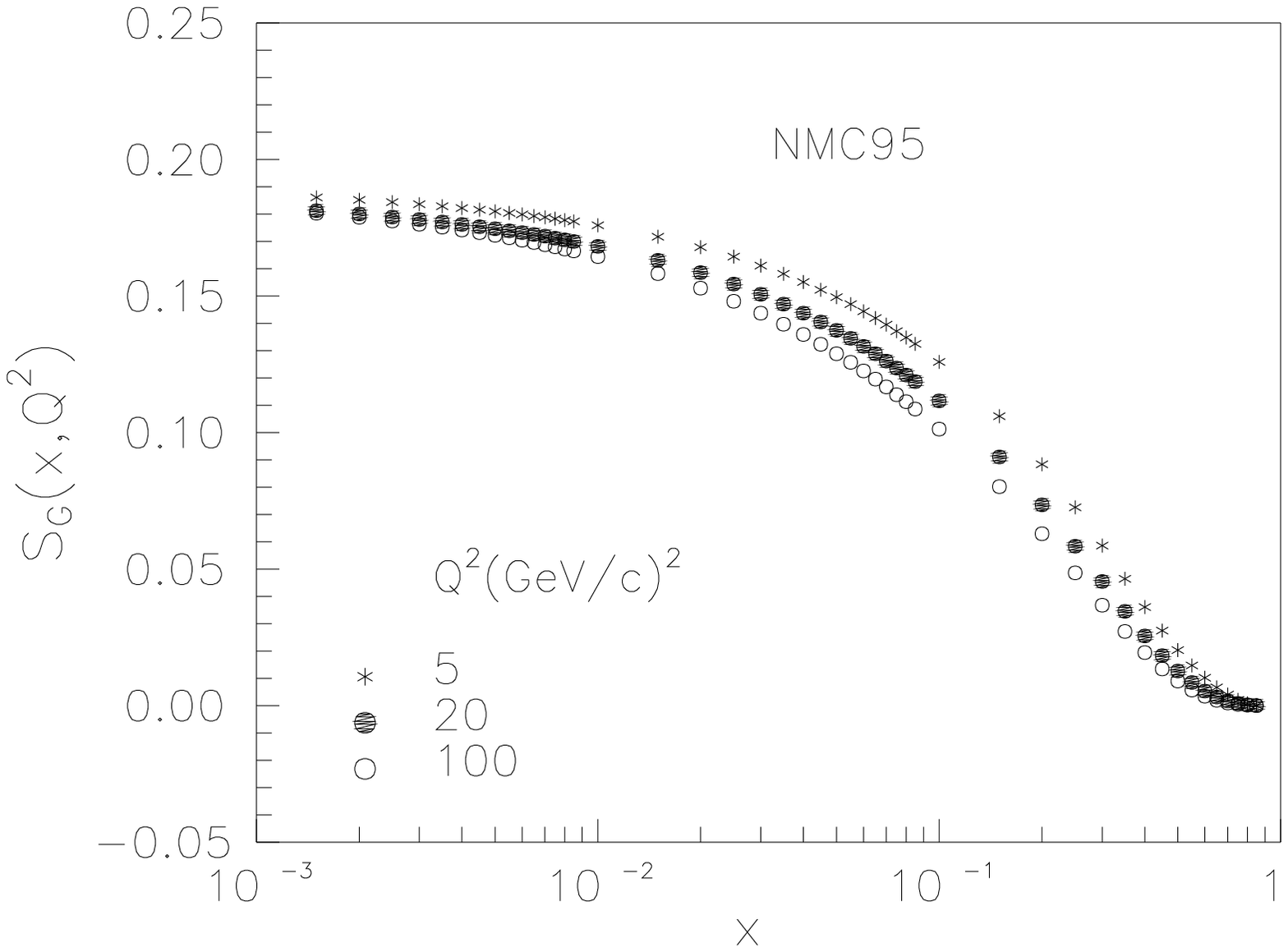}{}}
\parbox{7cm}{\epsfxsize=7.cm \epsfysize=7.cm \epsfbox[35 5 535 500]
{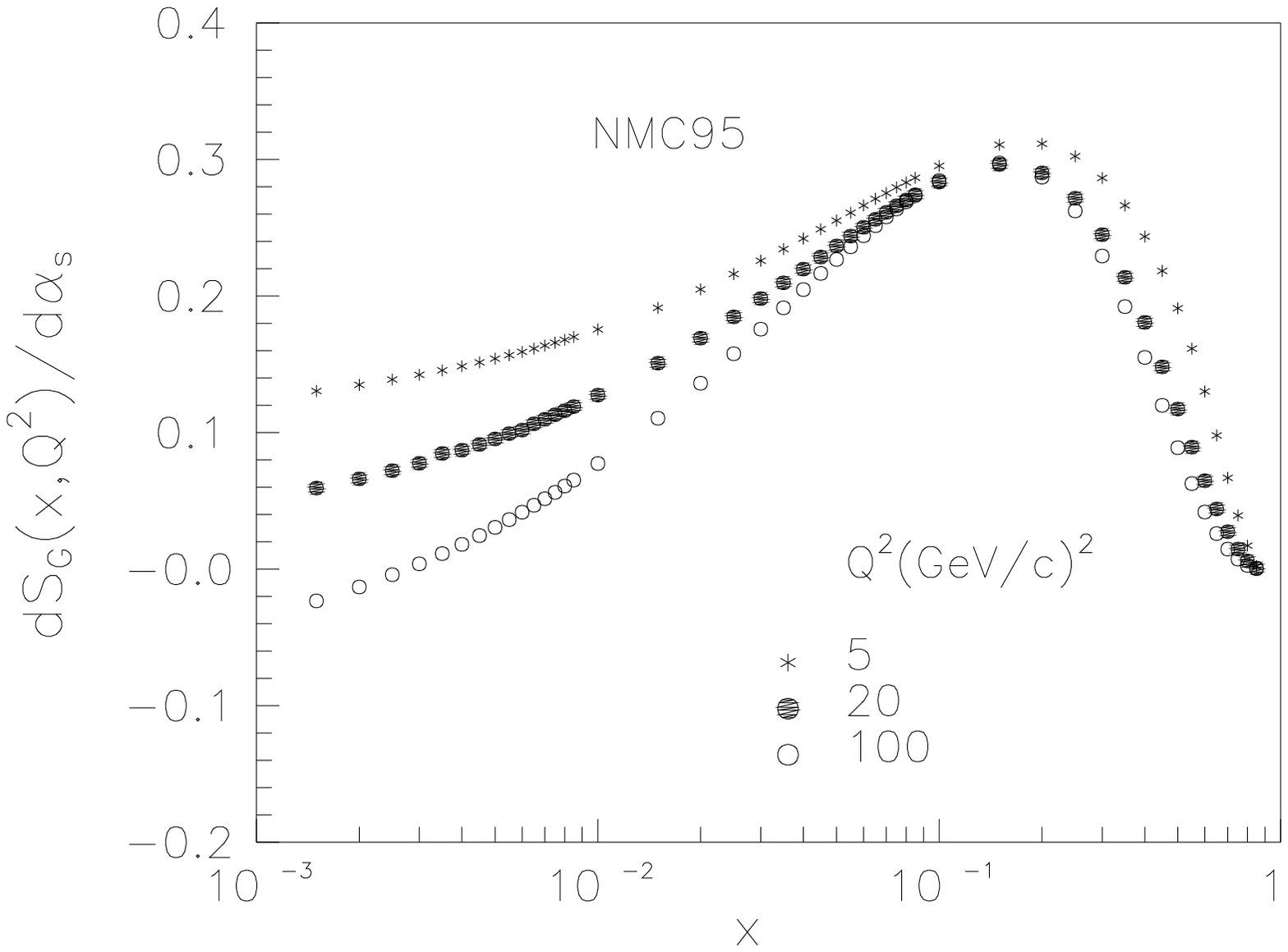}{}}
\end{center}

\begin{center}
\hspace*{0cm} c)\hspace*{7cm} d)
\end{center}

\newpage
\begin{center}
\vskip -1.cm
\parbox{7cm}{\epsfxsize=7.cm \epsfysize=7.cm \epsfbox[5 5 500 500]
{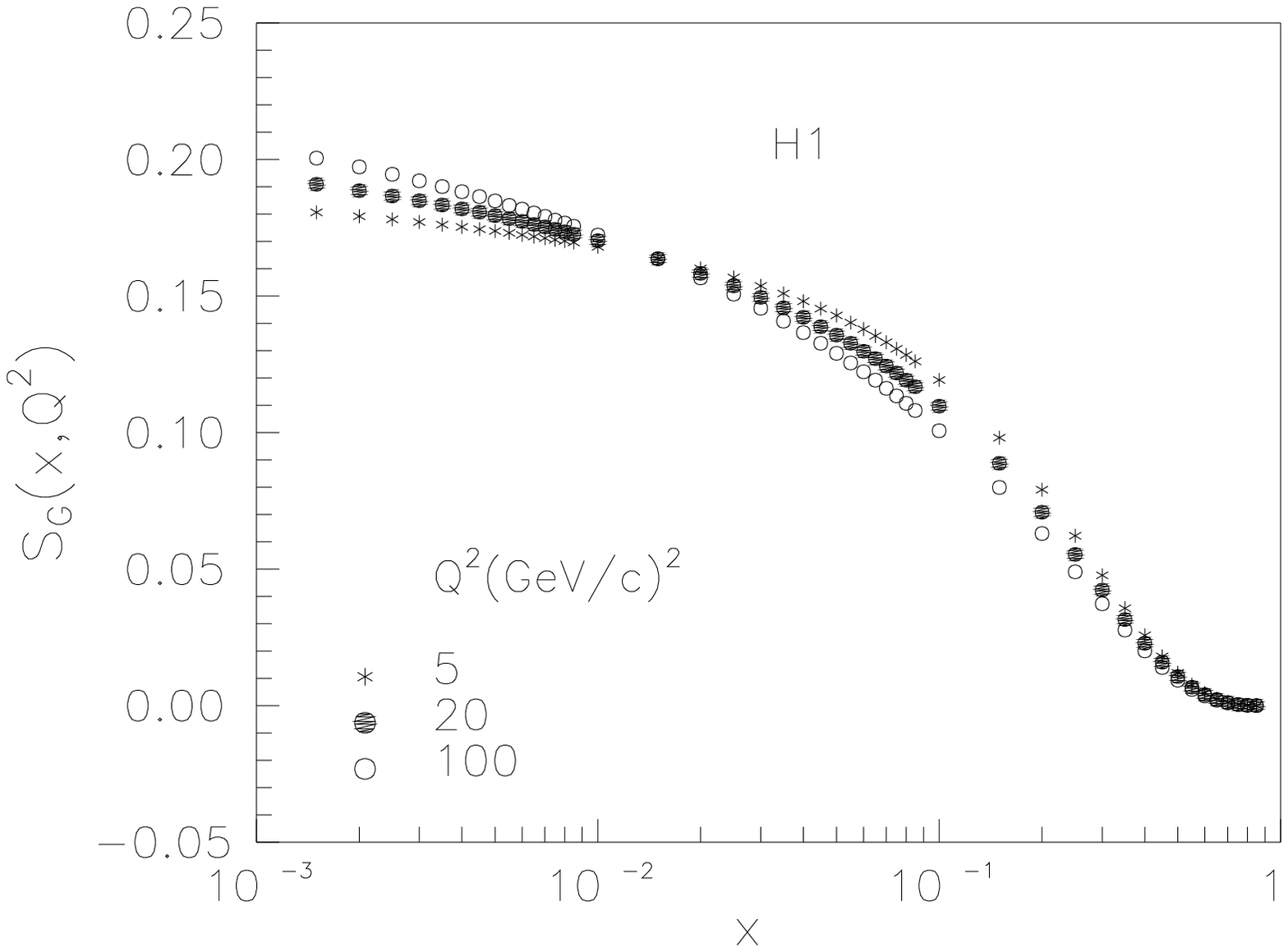}{}}
\parbox{7cm}{\epsfxsize=7.cm \epsfysize=7.cm \epsfbox[35 5 535 500]
{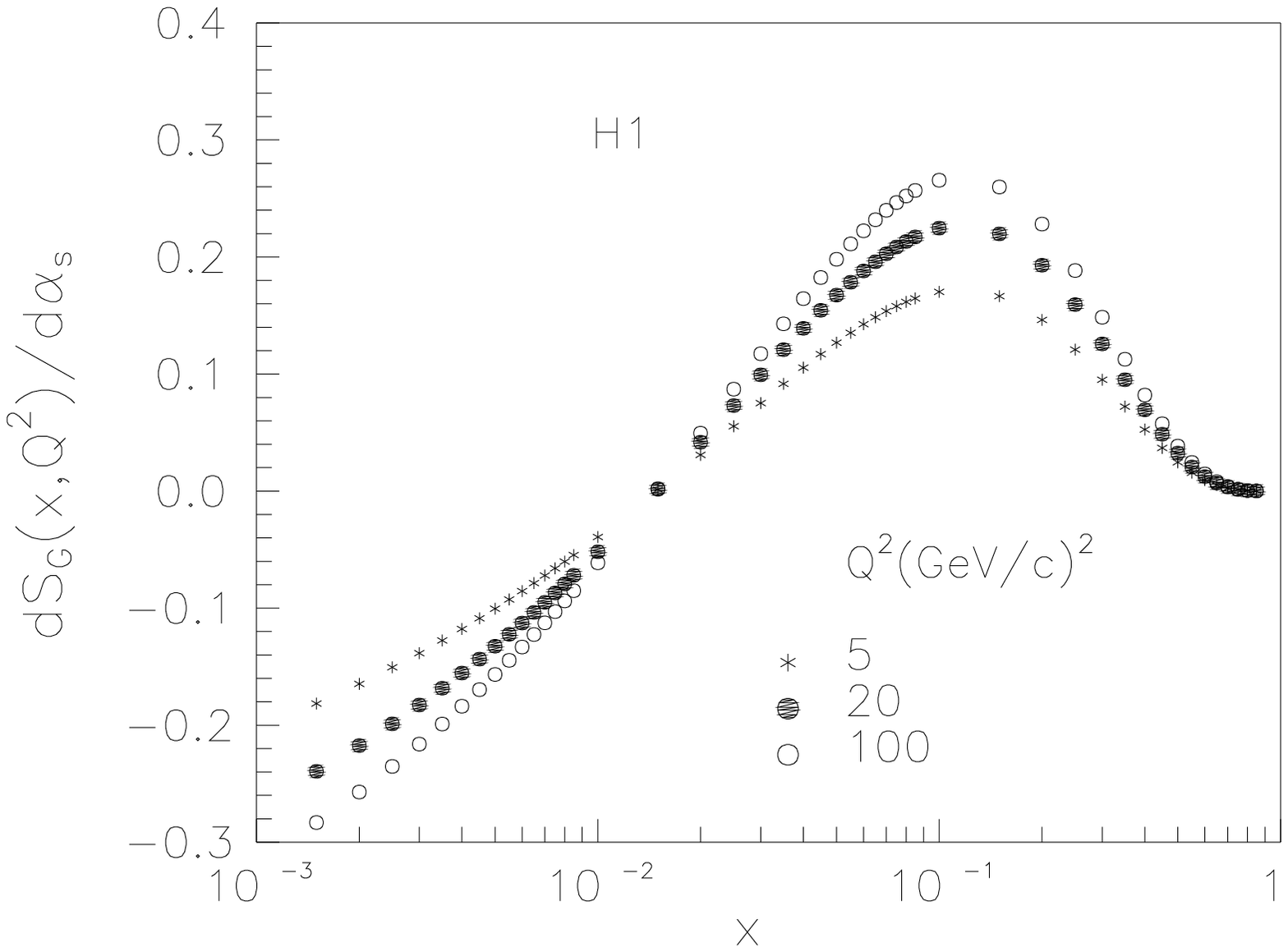}{}}
\end{center}

\begin{center}
\hspace*{0cm} e) \hspace*{7cm} f)
\end{center}

{\bf Figure 1.} The  Gottfried integral  $S_G(x,Q^2)$  and  the
derivative $dS_G(x,Q^2)/d\alpha_S$ as a function of $x$ and $Q^2$.

\begin{center}
\parbox{9cm}{\epsfxsize=9.cm \epsfysize=9.cm \epsfbox[5 5 500 500]
{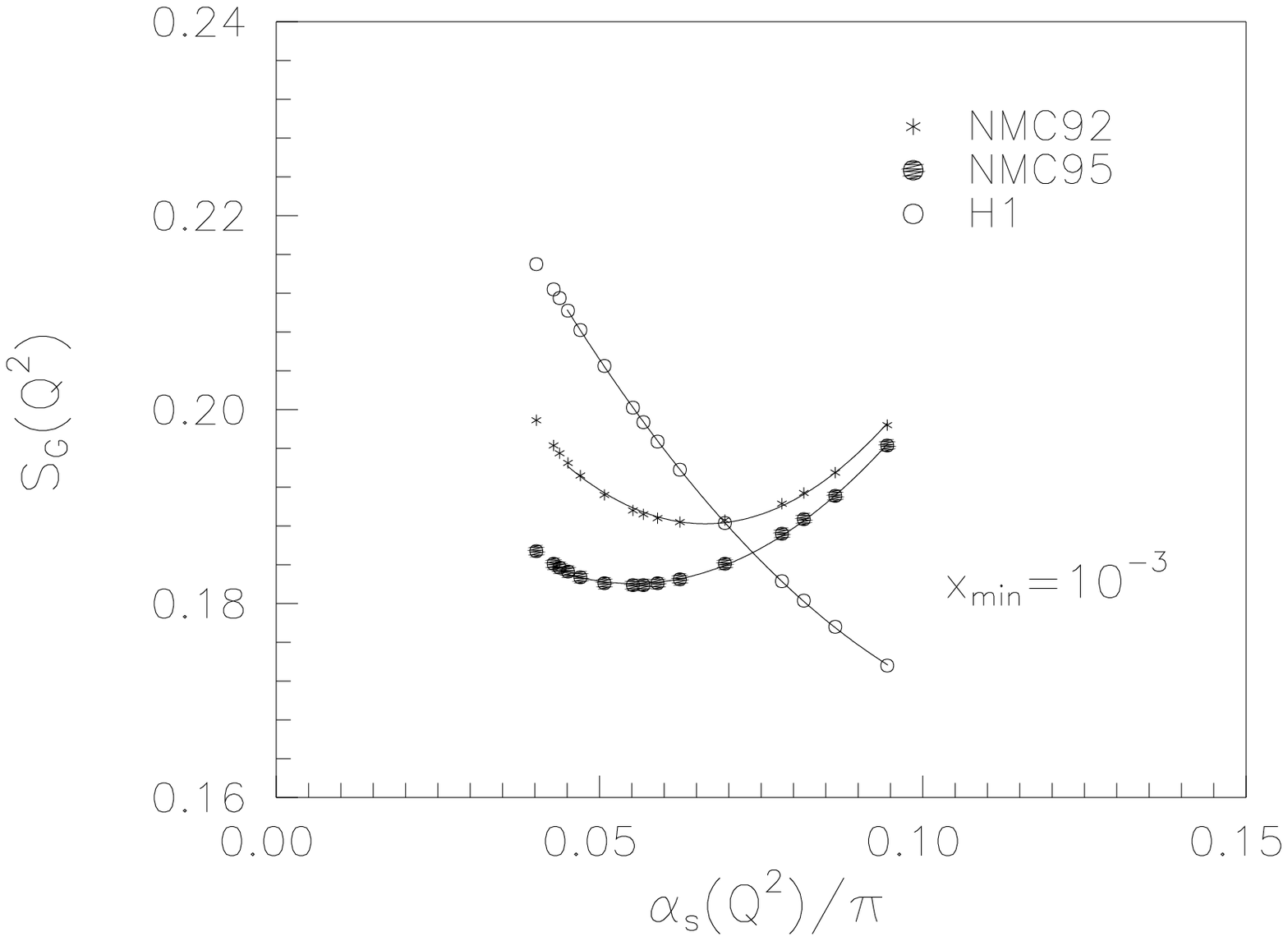}{}}
\end{center}

{\bf Figure 2.}
The Gottfried integral  $S_G(,Q^2)$ as a function of $\alpha_S$
 at  $x=10^{-3}$. The lines present the paraboloic fit  of $S_G(,Q^2)$
 for different parameterisations $F_2^p$.

\end{document}